\documentclass[12pt]{article} 
\usepackage{latexsym} 
\def\a{\alpha} 
\def\b{\beta} 
\def\g{\gamma}  
\def\d{\delta}  
\def\e{\eta} 
\def\G{\Gamma}  
  
\def\O{\Omega}
\def\m{\mu}  
\def\n{\nu}  
\def\r{\rho}  
 
\def\s{\sigma}

\def\e{\varepsilon} 
\def\cA{{\cal A}}
 
\def\pa{\partial}
 
\def\ll{\left} 
\def\rr{\right} 
\def\be{\begin{equation}}  
\def\ee{\end{equation}}  
\def\beq{\begin{eqnarray}}  
\def\eeq{\end{eqnarray}}  
\def\nn{\nonumber}

\def\cg{{\cal G}}

\def\cl{{\cal L}}

\def\E0{{E^{\left(0\right)}}}
\def\bmu{{\bar{\mu}}}
\newtheorem{definizione}{Definition}[section]
\newcommand{\bdefi}{\begin{definizione}}
\newcommand{\edefi}{\end{definizione}}
\makeatletter
\@addtoreset{equation}{section}
\makeatother
 
\begin{document} 
\begin{titlepage} 
\begin{flushright}  
hep-th/0012003 \\
ULB-TH-00/30\\ 
IHP-2000/06\\
\end{flushright}  
\vskip 2cm  
\begin{centering} 
{\huge {\bf An exotic theory of massless spin--two fields in three 
dimensions}} \\
\vspace{3cm} 
{\Large Nicolas Boulanger$^1$ and   Leonardo Gualtieri} \\ 
\vspace{1cm} 
Physique Th\'eorique et Math\'ematique,  Universit\'e Libre 
de Bruxelles,\\  
C.P. 231, B-1050, Bruxelles, Belgium      \\ 
Centre Emile Borel, Institut Henri Poincar\'e, 11 rue Pierre et Marie Curie,\\
F- 75231 Paris Cedex 05, France\\
\vspace{1cm}
{\tt\hskip 4cm nboulang@ulb.ac.be, lgualtie@ulb.ac.be} 
\vspace{2cm} 
\end{centering} 
\begin{abstract}
It is a general belief that the only possible way to consistently 
deform the Pauli--Fierz action, changing also the gauge algebra, 
is general relativity. 
Here we show that a different type of deformation exists in three 
dimensions if one allows for PT non--invariant terms. The new gauge 
algebra is different from that of diffeomorphisms. Furthermore, 
this deformation can be generalized to the case of a collection of 
massless spin--two fields. In this case it describes a 
consistent interaction among them.
\end{abstract} 
\vspace{2mm} \vfill \hrule width 3.cm
{\footnotesize
 $^1 $ \hskip 0.1cm 
''Chercheur F.R.I.A.'', Belgium.
\vspace{2mm}  \hrule width 3.cm}
\vfill           
\end{titlepage}
\section{Introduction} 
It is a general belief that the only possible gauge algebras for 
a theory of a massless spin--two field are the abelian 
and the diffeomorphism algebras.
However, if we allow for deformations that 
break PT invariance there is another possibility in three spacetime 
dimensions (and perhaps in five), as we show in this paper.
\par
The possible gauge algebras for massless spin--two fields 
have been recently studied in \cite{multigraviton}, in the 
context of an investigation on the problem of consistent couplings for a 
collection of massless spin--two fields, carried out by using 
BRST--based techniques (see \cite{MarcDefo1} and references therein).
The main result of \cite{multigraviton} was that theories involving 
different types of massless spin--two fields with non trivial, consistent, 
cross-interactions do not exist.  
This no-go theorem holds under the assumption that (i) the
Lagrangian contains no more than two derivatives of the massless spin-2 
fields 
$\{h_{\mu \nu}^a\}$ ($a = 1, \cdots, N$); (ii) the interactions can be
continuously switched on; (iii) the interactions are local and Poincar\'e 
invariant; (iv) in the limit of no interaction, the
action reduces to the sum of 
one  Pauli-Fierz action \cite{Fierz:1939ix}   
for each field $h_{\mu \nu}^a$, i.e. 
\beq 
\label{startingpoint} 
I_0[h_{\mu \nu}^a] &=& \sum_{a = 1}^N \int d^n x \ll[ 
-\frac{1}{2}\ll(\pa_{\m}{h^a}_{\n\r}\rr)\ll(\pa^{\m}{h^a}^{\n\r}\rr) 
+\ll(\pa_{\m}{h^a}^{\m}_{~\n}\rr)\ll(\pa_{\r}{h^a}^{\r\n}\rr)\rr.\nn\\ 
&&\ll.-\ll(\pa_{\n}{h^a}^{\m}_{~\m}\rr)\ll(\pa_{\r}{h^a}^{\r\n}\rr) 
+\frac{1}{2}\ll(\pa_{\m}{h^a}^{\n}_{~\n}\rr) 
\ll(\pa^{\m}{h^a}^{\r}_{~\r}\rr)\rr] 
\eeq 
(spacetime indices are raised and lowered with the flat Minkowskian metric
$\eta_{\mu \nu}$, for which we use a ``mostly plus" signature).
The free action (\ref{startingpoint}) 
is invariant under the linear gauge transformations, 
$\delta_\epsilon h^a_{\mu \nu} = \partial_\mu \epsilon_\nu^a 
+ \partial_\nu \epsilon_\mu^a $.
These transformations 
are abelian and irreducible.  The Pauli-Fierz action is in fact the 
linearized Einstein action and describes a massless spin-2 system. 
\par
To prove this statement, the techniques of \cite{MarcDefo1}
were used to find all the possible deformations of 
(\ref{startingpoint}) which deform the gauge algebra  (in the 
language of \cite{MarcDefo1}, the ones with $a_2\neq 0$) and satisfy 
the conditions (ii), (iii), and (iv).
Then, the deformations which modify the gauge 
transformation without changing the gauge algebra ($a_1\neq 0,~a_2=0$) 
and the ones that do not modify the gauge transformations 
($a_0\neq 0,~a_1=a_2=0$) were ruled out, by imposing the conditions (i), 
because all of them contain more than two derivatives. 
It was found that the only possible deformation of the abelian 
gauge algebra of (\ref{startingpoint}) is a direct sum of independent 
diffeomorphism algebras, and the corresponding action is a sum of 
independent Einstein--Hilbert (or possibly Pauli--Fierz) actions.
\par
However, the possibility of deformations which break PT--invariance 
was not considered in \cite{multigraviton}, also because they would
lead to interaction terms with three derivatives. As we will show in 
section 3, there are two such ''exotic'' deformations which 
solve the consistency equations at first order in the deformation 
parameter $\mu$
\be
I=I_0+\mu\int d^n\!x\,\,a_0+O(\mu^2)\,.
\ee
One is in $d=3$ spacetime dimensions
\be
a_0=\frac{1}{3}\e^{\a\b\g}\eta^{\r\s}\pa_{[\r}h^{b}_{\m]\a}
\pa^{[\m}h^{\n]a}_{\b}
\pa_{[\s}h^{c}_{\n]\g}a_{abc}\label{def3}
\ee
and one in $d=5$ spacetime dimensions
\be 
a_0=64\e^{\a\b\g\r\s}\pa_{[\r}h^{b}_{\m]\a}
\pa^{[\m}h^{\n]a}_{\b}
\pa_{[\s}h^{c}_{\n]\g}a_{abc}\,.\label{def5}
\ee
The corresponding gauge transformations are, respectively, 
\be
\d_{\epsilon} h^{\a\m a}=2\pa^{(\a}\epsilon^{\mu)a}+
\frac{\mu}{2}\e_{\b\g\d}
\ll(\pa^{[\b}h^{\a]\mu b}+\pa^{[\b}h^{\mu]\a b}\rr)\pa^{\g}\epsilon^{\d c}
a^a_{bc}+O(\mu^2)
\label{def3g}
\ee
and
\be
\delta_{\epsilon} h^{a\alpha\sigma}=2\pa^{(\a}\epsilon^{\s)a}
-2\mu(\varepsilon^{\alpha\beta\gamma\delta\rho}
\partial_{\beta}h^{b\,\s}_{\gamma}+
\varepsilon^{\s\beta\gamma\delta\rho}
\partial_{\beta}h^{b\,\a}_{\gamma})
\partial_{\delta}\epsilon^{c}_{\rho}a^a_{bc}
+O(\mu^2)\,.
\label{def5g}
\ee
These deformations involve one more derivative than the Einstein theory
(so, the coupling constants have different dimensions), 
and break PT--invariance. This is similar to what happens in the
Freedman--Townsend model \cite{FT}
for two--form fields in four dimensions, and in \cite{anco} for one--form  
fields in three dimensions, which have one more derivative than Yang--Mills
theory and break PT--invariance.
\par
In the three dimensional case (\ref{def3}), (\ref{def3g}) 
the spin--two fields take 
values in a commutative, symmetric (see \cite{multigraviton}), but not 
necessarily associative algebra $\cA$. In the five dimensional case 
(\ref{def5}), (\ref{def5g}) 
the algebra $\cA$ is anticommutative, and then there must 
be more than one field. These algebras are defined by means of the 
constants $a^a_{bc}$.
\par
The fact that (\ref{def3}), (\ref{def3g}) and (\ref{def5}), (\ref{def5g}) 
satisfy the consistency 
equations at first order in the deformation parameter does not guarantee 
a priori that they can be extended to 
deformations consistent at all orders.
In the case of (\ref{def3}), (\ref{def3g}), 
which arises in three spacetime dimensions, 
this extension is actually possible.
In fact, it is possible to write a complete expression, in first order 
formalism, for the action and the gauge symmetry. 
We derive it for the case of a single massless 
spin--two field and, then, for the 
case of a collection of massless spin--two fields. 
In order to find this action in first order formalism we consider the 
Chern--Simons formulation of gravity in three dimensions \cite{witten}, 
and change the $ISO(2,1)$ algebra to 
\beq
\ll[J_m,J_n\rr]&=&\varepsilon_{mnp}\ll(J^p+\mu P^p\rr)\nn\\
\ll[J_m,P_n\rr]&=&\varepsilon_{mnp}P^p\nn\\
\ll[P_m,P_n\rr]&=&0\,.
\label{newalgebra}
\eeq
Actually, this is only a change of basis, and then it still describes 
general relativity. However, if we ''switch off'' the Einstein 
interaction, 
that is, expand the fields in the dimensionful constant $\ell$ and send 
it to zero, we do not find general relativity, but a first order action 
with a new $\mu$--dependent gauge invariance. 
The second order formulation is a formal infinite series 
in $\mu$, and, up to $O(\mu)$, coincides 
with (\ref{def3}), (\ref{def3g}).
\par
We stress that such a theory does not admit a Chern--Simons
formulation (even if it can be obtained by taking a limit from a
Chern--Simons theory). So, it seems difficult to find a geometrical 
structure underlying this theory. 
Maybe this issue could be addressed by looking for something similar
to the actions studied in \cite{CCFM}.
\par
Generalizing to a collection of spin--two fields
\footnote{Collections of massless spin--two fields were investigated 
before in \cite{wald}.}, we find 
that the no--go theorem found in \cite{multigraviton} for consistent 
interactions does not apply. In fact, an essential ingredient for that 
theorem was the associativity of the algebra $\cA$ in which the fields 
take values, while in the deformation (\ref{def3}), (\ref{def3g}) 
the algebra $\cA$ is in general not associative.
\par
In the five dimensional case (\ref{def5}), (\ref{def5g}), 
it seems impossible to 
prove that the deformation is consistent to all orders working as in the 
case (\ref{def3}), (\ref{def3g}). 
Then, in principle, this deformation could be 
obstructed at some higher order. However, if it is consistent, it 
would be a very interesting theory, because in this case the algebra 
$\cA$ is anticommutative (and could be a Lie algebra). Furthermore, 
such a theory would be defined in $d=5$ spacetime dimensions, which 
constitutes an arena for several interesting recent developments of 
theoretical physics.
\par
As a remark, we consider also 
what happens when, starting from a single Pauli--Fierz action, 
we turn on both the PT--breaking deformation (that at first order 
in $\mu$ is (\ref{def3}), (\ref{def3g})) and general relativity.
The resulting theory is formally 
related by a field redefinition (or, more 
precisely, a BRST transformation) to 
the so called ''topological massive gravity'' 
\cite{DJT} found by Deser, Jackiw and Templeton. 
However, such a field redefinition can be defined only order by order in 
the deformation parameter, while non--perturbatively it is not 
invertible and changes the number of degrees of freedom.
\par
In section 2, we briefly recall the master--equation approach 
to the problem of consistent interactions. In section 3 we show 
that, if we do not require PT--invariance, there are new solutions of 
the problem studied in \cite{multigraviton}, at least at first 
order in the deformation parameter, in three and five spacetime 
dimensions.  In section 4 we find a complete expression, in first order 
formalism, for the action and the gauge symmetry corresponding to the 
deformation (\ref{def3}). The deformation of a single 
Pauli--Fierz action by both general relativity and (\ref{def3}) 
is discussed in appendix A.
\par
\section{Cohomological formulation} 
\par
A detailed exposition of the ideas of deformation theory
using the BRST cohomology techniques in Batalin-Vilkovisky 
formalism can be found for example in
\cite{MarcDefo1} together with useful references. 
Here we only briefly describe these features.
\par
\subsection{Gauge symmetries and master equation} 
\par
Let us consider an irreducible gauge theory with action $S[\Phi^i]$ 
whose gauge symmetries are given
by
\be
\d_{\epsilon}\Phi^i=R^{i}_{\a}(\Phi)\epsilon^{\a},
\label{gaugetransfo}
\ee
and gauge algebra
\be
R^j_{\a}\ll(\Phi\rr)\frac{\d R^i_{\b}\ll(\Phi\rr)}{\d \Phi^j} 
-R^j_{\b}\ll(\Phi\rr)\frac{\d R^i_{\a}\ll(\Phi\rr)}{\d \Phi^j} 
=C^{\gamma}_{\a\b}\ll(\Phi\rr)R^i_{\gamma}\ll(\Phi\rr)\, 
+ M_{\a\b}^{i j}\ll(\Phi\rr) \frac{\d S}{\d \Phi^j}. 
\label{gaugealg}
\ee
When $ M_{\a\b}^{i j}\neq 0$
\footnote{Here we are using the De Witt's condensed 
notation,  in which a summation over a repeated index 
implies also an integration. The $R^i_{\a}\ll(\Phi\rr)$ stand for 
$R^i_{\a}(x,x')$ and are combinations of the Dirac 
delta function $\delta(x,x')$ and some of its derivatives with 
coefficients 
that involve the fields and their derivatives, so that 
$R^i_{\a} \e^{\a} \equiv \int d^n x' R^i_{\a}(x,x') \e^{\a}(x')$
is a sum of integrals of $\e^{\a}$ and a finite number of its 
derivatives.},
the gauge transformations close only on shell.
The Noether identities read 
\be 
\frac{\d S}{\d \Phi^i} R^i_{\a} = 0. 
\label{Noether} 
\ee 
One can derive higher order 
identities from (\ref{gaugealg}) and (\ref{Noether}) 
by differentiating (\ref{gaugealg}) with respect to the fields.  
These identities, in turn, lead to further identities 
by a similar process. 
\par
It has been established in \cite{dwvh,bv1} that 
to every action $S$ one can associate a functional $W$ depending on 
the original fields $\Phi^i$ and on additional fields : the ghosts
$C^{\a}$ and the antifields $\Phi^{*}_i$, $C^*_{\a}$.
The fields $\Phi^i$ and $C^*_{\a}$ are bosonic while $\Phi^{*}_i$ and
$C^{\a}$ are fermionic.
\par
$W$ starts like 
\be
\label{writing}
W = S + \Phi^*_iR^i_{\a}C^{\a}+\frac{1}{2} 
\Phi^*_{\gamma}C^{\gamma}_{\a\b}C^{\a}C^{\b} 
+ \frac{1}{2} \Phi^*_i \Phi^*_j M_{\a\b}^{i j} C^\a C^\b  
+ \hbox{ ``more''}\,
\ee
(where ''more''  contains at least three ghosts) and fulfills 
what is called the {\it master equation}
\be
(W,W)=0\,.
\label{master}
\ee
The bracket $(.,.)$ (called {\it{antibracket}}) makes the fields 
and the antifields canonically conjugate to each other.
It is defined by 
\be 
(A,B) =  \frac{\d^R A}{\d \Phi^i} \frac{\d^L B}{\d \Phi^*_i} 
- \frac{\d^R A} {\d \Phi^*_i} \frac{\d^L B}{\d \Phi^i} 
+ \frac{\d^R A}{\d C^\a} \frac{\d^L B}{\d C^*_\a} 
- \frac{\d^R A}{\d C^*_\a} \frac{\d^L B}{\d C^\a}, 
\ee 
where the superscript $R$ ( $L$) denotes a right ( left) derivative,
respectively. The antibracket satisfies a graded Jacobi identity
(cfr \cite{BOOK} for all the information about BRST symmetry
and Batalin-Vilkovisky formalism). 
\par
The solution of the master
equation (\ref{master}) and the expression (\ref{writing}) for $W$ 
give us all the information about the gauge symmetries
of the action $S$ (gauge transformations (\ref{gaugetransfo}), 
gauge algebra (\ref{gaugealg}) ). 
The master equation is fulfilled as a consequence of 
the Noether identities (\ref{Noether}), of the gauge 
algebra (\ref{gaugealg}) and of all the higher order 
identities alluded to above that one can derive from them. 
Conversely, given some $W$, solution of (\ref{master}), one can recover 
the gauge-invariant action as the term independent of the ghosts 
in $W$, while the gauge transformations are defined by the terms 
linear in the antifields $\Phi^*_i$ and the structure functions 
appearing in the gauge algebra can be read off from the terms 
quadratic in the ghosts.  The Noether identities  
(\ref{Noether}) are fulfilled 
as a consequence of the master equation (the left-hand side of the 
Noether identities is the term linear in the ghosts in $(W,W)$; 
the gauge algebra (\ref{gaugealg}) is the next term in 
$(W,W) = 0$). 
In other words, there is complete equivalence between gauge invariance 
of $S$ and the existence of a solution $W$ of the master equation.
\par
Besides the ``fermionic'' grading given to the algebra $\cg$
of the dynamical variables, one has endowed this algebra with a 
$Z$-valued ``ghost grading'' called {\it{ghost number}} and an 
other $Z$-valued grading for 
the antifields called {\it{antighost number}}, or sometimes
{\it{antifield number}}.
\par
In Batalin-Vilkovisky formalism,
all the fields of linearized gravity in metric formulation
where
\be
g^a_{\m\n}=\eta_{\m\n}+\kappa^a h^a_{\m\n}
\label{metric}
\ee
are given by 
\begin{itemize} 
\item the fields $h^a_{\a\b}$, with ghost number zero and antifield number zero; 
\item the ghosts $C^a_{\a}$, with ghost number one and antifield number 
zero; 
\item the antifields $h^{* \a\b}_{a}$, with ghost  
number minus one and antifield  
number one; 
\item the antifields $C^{* \a}_{a}$, with ghost number minus  
two and antifield  
number two. 
\end{itemize} 
The action for a collection $\{h^a_{\m\n}\}$ of N non-interacting
$(a=1,\ldots,N)$ massless spin-2 fields writes \cite{Fierz:1939ix}
\beq 
\label{PFalgebradelta} 
I_0&=&\int d^nx~k_{ab}\ll[ 
-\frac{1}{2}\ll(\pa_{\m}{h^a}_{\n\r}\rr)\ll(\pa^{\m}h^{b\n\r }\rr) 
+\ll(\pa_{\m}h^{a \m}_{~\n}\rr)\ll(\pa_{\r}h^{b \r\n}\rr)\rr.\nn\\ 
&&\ll.-\ll(\pa_{\n}h^{a \m}_{~\m}\rr)\ll(\pa_{\r}h^{b \r\n}\rr) 
+\frac{1}{2}\ll(\pa_{\m}h^{a \n}_{~\n}\rr) 
\ll(\pa^{\m}h^{b \r}_{~\r}\rr)\rr]\,, 
\eeq
with a quadratic form $ k_{a b}$ defined by the kinetic terms.
In the way of writing the Pauli-Fierz action above,
$k_{ab}$ is simply equivalent, modulo field redefinitions, to the
Kronecker delta $\d_{ab}$. This is essential for the physical
consistency of the theory (absence of negative-energy 
excitations, or stability of the Minkowski vacuum). 
The gauge transformations  are
\be
\d_{\epsilon}h^a_{\a\b}=R^{a\,\g}_{(\-0\-)b,\a\b}
\epsilon^b_{\g}=\pa_{\a}\epsilon^a_{\b}+\pa_{\b}\epsilon^a_{\a}
\label{abeltransfo}
\ee 
where $\epsilon^a_{\a}$ are $n\times N$ arbitrary independent functions.
These transformations are abelian and irreducible.
The solution of the master equation for the free theory is
\be 
W_0 = I_0 + \int d^nx \, h^{* \a\b}_{a} (\partial_\a C^a_{\b} 
+ \partial_\b C^a_\a). 
\ee 
\par 
We define the BRST operator and its action on a functional $A$
by
\be
sA=(W_0,A),
\ee
that is, as a canonical transformation in some extended phase space
\footnote{Namely, $W_0$ is the generator of the ``canonical'' 
transformations.}.
This enables us to get the BRST differential $s$ of the free 
theory as  
\be 
s = \delta + \gamma 
\ee 
where the action of $\gamma$ and $\delta$ on the variables is  
zero except 
\beq 
\g h^a_{\a\b}&=&2\pa_{(\a}C^a_{\b)} \label{trivialdc} \label{defg}\\ 
\d h_a^{*\a\b}&=&\frac{\d I_0}{\d h^a_{\a\b}}\label{defd1}\\ 
\d C_a^{*\a}&=&-2\pa_{\b}h_a^{*\b\a} \,. \label{defd2} 
\eeq
Note in particular that $ \g C^a_\a = \d C^a_\a =0$.
\par
The nilpotency of the differential $s$ follows from the graded Jacobi
identity for the antibracket and from the fact that $W_0$ satisfies 
the free master equation $sW_0=(W_0,W_0)=0$. 
The decomposition of $s$ into $\delta$ plus $\gamma$ is 
dictated by the antifield number: $\delta$ decreases the 
antifield number by one unit, while $\gamma$ leaves it unchanged. 
Combining this property with $s^2 =0$, one concludes that
\be 
\delta^2 = 0, \; \delta \gamma + \gamma \delta = 0, \; 
\gamma^2 = 0. 
\ee 
The differential $\g$ is the longitudinal derivative along the 
gauge orbits, while $\d$ enables us to implement the field equations 
in a cohomological construction.
\par
\subsection{Consistent deformations as a cohomological problem}
\par
We would like now to deform the free action $I_0$ by adding to it 
interaction terms 
\be
I_0 \rightarrow I=I_0+\lambda I_1+\lambda^2I_2+\ldots
\label{defoact}
\ee
We associate to $S$ the functional $W$
\be 
\label{exprW}
W = W_0 + \lambda W_1 + \lambda^2 W_2 + O(\lambda^3)\,,
\ee 
where $W_0$ is the solution of the master equation for the 
free theory. $W$ also has to fulfill the master equation
\be
\label{masterW}
(W,W)=0\,,
\ee
in order for the deformation to be consistent.
It is worth noting that in this way, we find deformations with
the same number of physical degrees of freedom as the original theory 
(and also the same number of independent
gauge symmetries, reducibility identities, $\ldots$).
\par
So, the problem of deforming an action consistently turns out to be 
equivalent to the problem of deforming the solution $W_0$ of the 
master equation $(W_0,W_0)=0$ into a solution $W$ of
the deformed master equation $(W,W)=0$. 
We can treat this last problem perturbatively in power of the 
deformation parameter $\lambda$. 
Then we try to construct the deformations order by order in $\lambda$.
\par
Substituting (\ref{exprW}) in (\ref{masterW}) yields, 
up to order $\lambda^2$,
\begin{eqnarray} 
O(\lambda^0): & & (W_0,W_0) = 0 \label{key1}\\ 
O(\lambda^1): & & (W_0,W_1) = 0 \label{key2}\\ 
O(\lambda^2): & & (W_0,W_2) = - \frac{1}{2} (W_1,W_1). \label{key3} 
\end{eqnarray}
The first equation is fulfilled by assumption since the starting point 
defines a consistent theory. Then, the solutions of equation (\ref{key2})
give the deformations up to order $\lambda$. However, these deformations 
are actually consistent only if they fulfill 
equation (\ref{key3}) and the equations at higher order in $\lambda$.
This was proved, for example, in the case of the deformation studied 
in \cite{multigraviton}. Alternatively, one could take a deformation 
at order $\lambda$, solution of (\ref{key2}), and search for a consistent theory 
that, at first order in $\lambda$, coincides with that 
deformation. This is the approach we follow in this paper.
\par
Equation (\ref{key2}) can be rewritten as 
\be
s\,W_1=0\,.
\ee
On the other hand, it can be shown \cite{MarcDefo1} 
that solutions of the form
\be
W_1=s\,\Lambda
\ee
correspond to trivial deformations, e.g. field redefinitions. 
So, the first-order non-trivial deformations of $W_0$
are elements of the BRST cohomology group (in ghost number 0)
$H_0(s)$ \cite{Barnich:1993}. 
Because the equation $s \int a = 0$ is equivalent to $sa + dm = 0$ 
for some $m$, and $\int a = s \int b$ is equivalent to $a = s b + dn$ 
for some $n$, one denotes the corresponding cohomological 
group by $H^{0,n}(s \vert d)$ where the integrand
$a$, $b$, $m$, $n$ are {\em local forms}, that is, differential forms 
with local functions as coefficients. 
{\em Local functions} depend polynomially on the fields
(including the ghosts and the antifields) and their derivatives 
up to a finite order
(in such a way that we work with functions over a finite-dimensional 
vectorial space, the so-called jet space).
\par
To compute the consistent, first order deformations, i.e., 
$H^{0,n}(s \vert d)$, one needs 
$H(\gamma)$ and $H(\delta \vert d)$.
For this computation we refer to \cite{multigraviton}
\par
\section{Cohomological deformations in 3 and 5 dimensions}
\par
From the computation of $H^{0,n}(s \vert d)$ (see \cite{multigraviton}) 
one finds that $W_1$ has, 
modulo BRST transformations, no term 
with antighost number greater than 2:
\be
W_1=\int (a_0+a_1+a_2)\,.
\ee
We recall that the antighost number zero term $a_0$ is the deformation of 
the action, the antighost number one term $a_1$ gives the deformation of 
the gauge transformations, and the antighost number two term $a_2$ gives
the deformation of the gauge algebra.
In this paper we are considering the deformations which change the gauge 
algebra, and have then $a_2\neq 0$. 
\par
From the study of $H^{0,n}(s \vert d)$ it emerges (see 
\cite{multigraviton} and \cite{BBH2}) that 
$a_0$, $a_1$ and $a_2$ have to satisfy the following equations
\footnote{As often in the sequel, we shall 
switch back and forth between a form and its dual without 
changing the notation when no confusion can arise.  So the 
same equation for $a$ is sometimes 
written as $sa + db = 0$ and sometimes written as $sa + \partial_\m 
b^\m = 0$.}:
\beq
\g a_0+\d a_1 &=& \pa_{\mu}j^{\mu}_0\label{g0d1}\\
\g a_1+\d a_2 &=& \pa_{\mu}j^{\mu}_1\label{g1d2}\\
\g a_2 &=& 0\,.\label{g2}
\eeq
The equation (\ref{g2}) determines $a_2$ modulo $\g$--exact terms, 
corresponding to BRST transformations, that is, $a_2\in H_2(\g)$. 
But a necessary condition for $a_2$ to be consistent is that 
(\ref{g1d2}) also has solutions, namely, $a_2\in H_2^n(\d\vert d)$.
It can be shown that this 
imposes to $a_2$ the following necessary condition:
it is linear in the antighosts $C^{*\m}$, while 
it contains no quadratic part in the antifields 
$h^*_{\m\n}$ (then, as one can see by comparison with the explicit 
form of $W$, the algebra closes off shell).
Then, the fact that $W$ has ghost number zero implies that 
$a_2$ is quadratic in the ghost or their derivatives. 
Furthermore, it can be shown that $a_2$ cannot depend on 
$h_{\a\b}$. Then, the antighost and the ghost have to be 
contracted among themselves, or with a Poincar\'e invariant tensor, 
that is, with $\e_{\a_1\dots\a_n}$. 
Terms with two derivatives on one ghost are excluded, because
\be
\pa_{\a\b}C^a_{\g}=\g\ll(\G^a_{\a\b\g}\rr)
\ee
with
\be
\G^a_{\a\b\g}\equiv\frac{1}{2}(\pa_{\a}h^a_{\b\g}+\pa_{\b}h^a_{\a\g}-
\pa_{\g}h^a_{\a\b})\,,
\ee
and $a_2\in H(\g)$.
It follows that the only few possibilities are 
\beq
a_2&=&-C^{*\b}_aC^{b\a}\pa_{[\a}C^c_{\b]}a^a_{bc}
~~{\mbox{in any dimension}}
\label{firsta2}
\\
a_2&=&\e^{\a\b\g}C^*_{a\a}C^b_{\b}C^c_{\g}a^a_{bc}~~~~ {\rm{and}}~~~~
a_2=\frac{1}{4}
\e_{\b\g\d}C^*_{a\a}\pa^{[\a}C^{b\b ]}\pa^{[\g}C^{b\d]}a^a_{bc}
 ~~ {\rm{in}}~~d=3~~;
\label{d3a2}
\\
a_2&=&\e_{\m\a\b\g}C^{*\m}_a\pa^{[\a}C^{b\b ]}C^{c\g}a^a_{bc}
~~{\rm{in}}~~d=4~~
{\rm{and}}
\label{d4a2}
\\
a_2&=&\e^{\a\b\g\d\r}C^*_{a\a}\pa_{\b}C_{\g}^b\pa_{\d}C_{\r}^ca^a_{bc}
~~{\rm{in}}~~d=5\,.
\label{d5a2}
\eeq
where $a^a_{bc}$ are constants, which define an algebraic structure 
on the space in which the fields take values. 
\par
Among these, only the deformation (\ref{firsta2}) was considered in 
\cite{multigraviton}, because the others, built up with the Levi--Civita 
tensor, break PT invariance, and because the corresponding  
deformations of the action (when (\ref{g0d1}), (\ref{g1d2}) are
satisfied) contain more than two derivatives.
However, as we said, it is an interesting issue to study 
these deformations. Then, in the following, we will consider the cases 
(\ref{d3a2}), (\ref{d4a2}), (\ref{d5a2}).
\par
\subsection{The three dimensional case}
\par
The algebra deformation
\be
a_2=\e^{\a\b\g}C^*_{a\a}C^b_{\b}C^c_{\g}a^a_{bc},
\ee
is not consistent because (\ref{g1d2}) has no solution.
In fact, modulo total derivatives,
\be
\delta
a_2=-4h_{a\alpha}^{*~~\sigma}\partial_{[\sigma}C^b_{\beta]}C^c_{\gamma}
\epsilon^{\alpha\beta\gamma}a^a_{bc}
\ee
is a non--trivial element of $H(\g\vert d)$.
\par
Let us consider the deformation 
\be
\label{a2d3}
a_2=\frac{1}{4}\e_{\b\g\d}C^*_{a\a}\pa^{[\a}C^{\b]b}\pa^{[\g}C^{\d]c}a^a_{bc}.
\ee
Equation (\ref{g1d2}) gives
\be
\d a_2 = \pa_{\m}j_1^{\m}+\g \left[\frac{1}{2}\e^{\b\g\d}h^*_{\m\a a}
\left( \pa^{[\a}h^{\b ]\m b}\pa^{\g}C^{\d c}-
       \pa^{[\a}C^{\b ] b}\pa^{\g}h^{\d\m c}\right)a^a_{bc}
       \right]\,.
\ee
It admits a solution, which is (modulo $\g$--exact terms) 
\be
a_1=-\e^{\b\g\d}h^*_{\m\a a}\pa^{[\a}h^{\b ]\m b}
\pa^{\g}C^{\d c}a^a_{(bc)}\,,
\label{a1d3}
\ee
provided 
\be
a^a_{bc}=a^a_{(bc)}\,.
\ee
The deformation (\ref{a1d3}) corresponds to the following gauge 
transformation:
\be
 \d_{\epsilon} h^{\a\m a}=2\pa^{(\a}\epsilon^{\mu)a}+
\frac{\mu}{2}\e_{\b\g\d}
\ll(\pa^{[\b}h^{\a]\mu b}+\pa^{[\b}h^{\mu]\a b}\rr)\pa^{\g}\epsilon^{\d c}
a^a_{(bc)}+O(\mu^2)\,.
\ee
We stress that (\ref{a1d3}) is defined modulo a solution of the homogeneous
equation $\g a_1 + \pa_{\mu}k^{\mu}_1 =0$, which do not deform the algebra.
For a discussion about this ambiguity, see \cite{multigraviton}. 
\par
Equation (\ref{g0d1}) 
admits a solution if and only if
\be
a_{abc}=a_{(abc)}
\ee
where $a_{abc}\equiv\d_{ad}a^d_{cb}$.
The solution (modulo $\gamma$--exact terms and total derivatives), 
namely, the deformation of the lagrangian, is
\be
\label{a0d3}
a_0=\frac{1}{3}\e^{\a\b\g}\eta^{\r\s}\pa_{[\r}h^{b}_{\m]\a}
\pa^{[\m}h^{\n]a}_{\b}
\pa_{[\s}h^{c}_{\n]\g}a_{abc}\,.
\ee
\par
The consistency conditions at second order in the perturbation is
\be
\label{11s2}
(W_1,W_1)=-2sW_2\,.
\ee
This condition, at antighost number two, is always satisfied:
\be
\label{3a22}
\left(a_2,a_2\right)=\g\left(\frac{1}{2}
\e_{\b\g\s}\e_{\rho\mu\nu}C^*_{\a a}\pa^{[\a}C^{\b]b}\pa^{[\s}C^{\rho]e}
\pa^{\mu}h^{\nu\g c}~a^a_{bd}a^d_{ec}\right)\,.
\ee
Notice that in the case of the deformation studied in \cite{multigraviton}, 
at this stage the associativity condition arise, while 
in our case no condition is required for $(a_2,a_2)$ to be $\g$--exact.
\par
Instead of verifying the rest of (\ref{11s2}) and the higher order conditions, 
we will provide in section 4 a shortcut to prove that this deformation is consistent,
namely, by constructing its complete expression.
\par
\subsection{The four dimensional case}
\par
The deformation
\be
a_2=\e_{\m\a\b\g}C_a^{*\m}\pa^{[\a}C^{\b ]b}C^{\g c}a^a_{bc}
\ee
is not consistent. In fact, trying to solve (\ref{g1d2})
one finds that $\d a_2$ is a non--trivial element of $H(\g\vert d)$, and 
there is no way to get rid of it by imposing conditions
on the coefficients $a^a_{bc}$. 
\par
\subsection{The five dimensional case}
\par
The deformation
\be
a_2 = \varepsilon^{\alpha\beta\gamma\delta\rho}C^{*}_{a\alpha}\partial_{\beta}C^{b}_{\gamma}
\partial_{\delta}C^{c}_{\rho}a^a_{bc}\,.\label{a2d5}
\ee
is not vanishing only if
\be
\label{antisymmetry}
a^a_{bc}=a^a_{[bc]}\,.
\ee
So this deformation can occur only if there is more than one 
spin--two field. The solution of the equation
\be
\delta a_2+\gamma a_1=\partial^{\mu}j_{1\mu}
\ee
(modulo deformation with $a_2=0$, see \cite{multigraviton})
is
\beq
a_1&=&-4\varepsilon^{\alpha\beta\gamma\delta\rho}h^{*~\sigma}_{a\alpha}\partial_{\beta}
    h^{b}_{\gamma \sigma} \partial_{\delta}C^{c}_{\rho}a^a_{bc}=\nn\\
&=&4\varepsilon^{\alpha\beta\gamma\delta\rho}h^{*~\sigma}_{a\alpha}
    \Gamma^{b}_{\beta\gamma \sigma} \partial_{\delta}C^{c}_{\rho}a^a_{bc}\,.
\eeq
The corresponding gauge transformation is
\be
\delta_{\epsilon} h^{a\alpha\sigma}=2\pa^{(\a}\epsilon^{\s)a}
-2\mu(\varepsilon^{\alpha\beta\gamma\delta\rho}
\partial_{\beta}h^{b\,\s}_{\gamma}+
\varepsilon^{\s\beta\gamma\delta\rho}
\partial_{\beta}h^{b\,\a}_{\gamma})
\partial_{\delta}\epsilon^{c}_{\rho}a^a_{bc}
+O(\mu^2)\,.
\ee
Equation
\be
\delta a_1+\gamma a_0=\partial^{\mu}j_{0\mu}
\ee
has solution if and only if
\be
a_{abc}=a_{[abc]}\,.\label{antisym}
\ee
The solution, which is the deformation of the Pauli Fierz lagrangian at the
first order in the perturbation, is
\be
a_0=64\varepsilon^{\alpha\beta\gamma\delta\rho}\partial_{[\alpha}h^{a}_{\xi]\beta}
     \partial^{[\xi}h^{(b)\sigma]}_{~~\gamma}\partial_{[\sigma}h^{c}_{\rho]\delta}a_{abc}\,.
\ee
Condition (\ref{antisymmetry}) suggests that maybe this theory describes an
interaction between Lie algebra--valued spin two fields. In this case, the 
Jacobi identity should arise from the
consistency conditions at second order in the perturbation, that is
\be
(W_1,W_1)=-2sW_2
\,.
\ee
This condition, at antighost number two, is always satisfied:
\be
\label{a22}
\left(a_2,a_2\right)=-
4\partial^{\eta}C^{*\lambda}_d\partial^{\mu}C^{f\nu}\partial_{\beta}C^b_{\gamma}\partial_{\delta}C^c_{\rho}
\varepsilon_{\lambda\mu\nu\rho\alpha}\varepsilon^{\alpha\beta\gamma\delta\rho}a^a_{bc}a^d_{af}
\ee
which is $\gamma$--exact, because integrating by parts one 
find all terms with second derivatives of ghosts. 
\par
We do not know more about this theory, in particular we do not know if the 
consistency conditions
at the higher orders are satisfied. In the following, 
we will consider only the deformation  (\ref{a0d3}), (\ref{a1d3}), 
(\ref{a2d3}), defined in three spacetime dimensions.
\par
\section{First order formulation of the three dimensional theory}
\par
In the previous sections 
we have proved the consistency of the deformation corresponding to 
(\ref{a0d3}) only up to first order. It is possible to show
that this deformation is consistent at all orders, and to find its
complete expression, by turning to ''first order formulation''.
We consider first the case of a single massless spin--two 
field. We find the action
\be
I^{EX}=\int \ll(\frac{1}{2}
\e_{mnp}h^m\wedge d\Omega^{np}+\frac{1}{2}
\e_{mnp}\wedge\E0^m\wedge
\Omega^n_{\,\,q}\wedge\Omega^{qp}+\frac{1}{3}\bmu\Omega_p^{\,\,m}\wedge
\Omega_{mn}\wedge\Omega^{np}\rr)\,,
\ee
which is invariant under the gauge transformation
\beq
\d h^m&=&2d\epsilon^m-2\E0_n\s^{mn}-
\bmu\e^{npq}\Omega_{mn}\s_{pq}\nn\\
\d\Omega_{mn}&=&d\s_{mn}\,,
\eeq
as can be checked directly.
Generalizing to a collection of such fields, the action is
\beq
I^{EX}&=&\int \ll(\frac{1}{2}\d_{ab}\ll(
\e_{mnp}h^{m\vert a}\wedge d\Omega^{np\vert b}
+\e_{mnp}\E0^m\wedge
\Omega^{n\vert a}_{\,\,q}\wedge\Omega^{qp\vert b}\rr)+\rr.\nn\\
&&\,\,\,\,\ll.+\frac{1}{3}\bmu_{abc}\Omega_p^{\,\,m\vert a}\wedge
\Omega_{mn}^b\wedge\Omega^{np\vert c}\rr)\,,
\eeq
invariant under the gauge transformations
\beq
\d h^{m\vert a}&=&2d\epsilon^{m\vert a}-2\E0_n\s^{mn\vert b}-
\bmu^a_{\,\,bc}\e^{npq}\Omega_{mn}^b\s_{pq}^c\nn\\
\d\Omega_{mn}^a&=&d\s_{mn}^a\,.
\eeq
\par
\subsection{Chern--Simons Gravity}
\par
As shown by E. Witten \cite{witten}, gravity in three dimensions can be
reformulated as a Chern-Simons gauge theory.
Let us briefly recall this formulation. 
\par
The kinetic term of the Chern--Simons action,
\be
d_{mn}A^m\wedge dA^n
\ee
(with $A$ Lie algebra valued gauge field) can exist only if the Lie 
algebra admits a non degenerate invariant metric $d_{mn}$. 
For a semisimple group the Killing metric is non degenerate, so a 
Chern--Simons formulation is allowed. On the contrary, in general 
one cannot construct a non--degenerate metric for a Poincar\'e 
group. However, there is an exception:
there exists a non degenerate metric for $ISO(2,1)$, corresponding 
to the invariant bilinear $W=\varepsilon_{mnp}P^mJ^{np}$.
If we replace the Lorentz generators $J^{mn}$ ($m,n=0,1,2$) with
\be
J^m=\frac{1}{2}\varepsilon^{mnp}J_{np}
\ee
we can write this metric as
\be
\label{CSmetric}
\langle J_m,P_n\rangle=\delta_{mn},~~~\langle J_m,J_n\rangle=
\langle P_m,P_n\rangle=0\,.
\ee
The commutation relations of $ISO(2,1)$ take the form
\beq
\ll[J_m,J_n\rr]&=&\varepsilon_{mnp}J^p\nn\\
\ll[J_m,P_n\rr]&=&\varepsilon_{mnp}P^p\nn\\
\ll[P_m,P_n\rr]&=&0\,.
\label{einalgebra}
\eeq
We can then construct a gauge theory for the group $ISO(2,1)$, by taking 
as gauge field the Lie--algebra valued one--form
\be
A=e^mP_m+\omega^mJ_m\,.
\ee
The corresponding gauge transformations are equivalent 
on--shell to diffeomorphisms, and the Chern--Simons action is
\be
\label{EH}
I^{CS}=2\int \ll(e^m\wedge\ll(d\omega_m+\frac{1}{2}\varepsilon_{mnp}
\omega^n\wedge\omega^p\rr)\rr)
\ee
which is the Einstein--Hilbert action in first order formalism.
\par
Let us consider the algebra
\beq
\ll[J_m,J_n\rr]&=&\varepsilon_{mnp}\ll(J^p+\mu P^p\rr)\nn\\
\ll[J_m,P_n\rr]&=&\varepsilon_{mnp}P^p\nn\\
\ll[P_m,P_n\rr]&=&0\,.
\label{defalgebra}
\eeq
The corresponding Chern--Simons action 
\be
\label{defCSaction}
I^{'CS}=\int \ll(2e_m\wedge\ll(d\omega^m+\frac{1}{2}\varepsilon^{mnp}
\omega_n\wedge\omega_p\rr)+\frac{1}{3}\mu\varepsilon^{mnp}\omega_m
\wedge\omega_n\wedge\omega_p\rr)
\ee
is invariant under the gauge transformations
\beq
\d e^m&=&d\epsilon^m-\varepsilon^{mnp}e_n\s_p+\varepsilon^{mnp}\omega_n
\epsilon_p+\mu\e^{mnp}\omega_n\s_p\nn\\
\d\omega^m&=&d\s^m+\e^{mnp}\omega_n\s_p\,.
\label{gaugeEin}
\eeq
The algebra (\ref{defalgebra}) is not a true deformation of 
$ISO(2,1)$. In fact, by a redefinition of the generators 
\be
\label{JmP}
J'_m=J_m+\mu P_m
\ee
(\ref{einalgebra}) becomes (\ref{defalgebra}).
\par
As a consequence, the action (\ref{defCSaction}) should describe 
the Einstein theory, and it actually occurs, at least at a classical 
level. If we perform the change of variable corresponding to (\ref{JmP}),
\be
e_m'=e_m+\mu \omega_m
\ee
the action (\ref{defCSaction}) 
becomes the sum of the Einstein--Hilbert action and the 
Chern--Simons action for $SO(2,1)$
\be
I^{'CS}=\int \ll( e^{'m}\wedge(2d\omega_m+\e_{mnp}\omega^n
\wedge\omega^p)
-\mu\omega^m\wedge(2d\omega_m+\frac{2}{3}\e_{mnp}\omega^n
\wedge\omega^p)
\rr)
\label{defCSaction2}
\ee
whose field equations are
\beq
2d\omega_m+\e_{mnp}\omega^n\wedge\omega^p&=&0\nn\\
2de_m'+\e_{mnp}e^{'n}\wedge\omega^p-2\mu\left(
2d\omega_m+\e_{mnp}\omega^n\wedge\omega^p\right)&=&0\,.
\eeq
This action doesn't admit a second order formulation, because 
the $\omega_m$ fields are no longer auxiliary fields.
However, the space of its solutions
coincides with the space of the solutions of the Einstein--Hilbert action.
\par
\subsection{The $\ell\rightarrow 0$ limit}
\par
Even if the first order actions (\ref{EH}), (\ref{defCSaction}) 
describe (at least at a classical level) the same theory, 
turning off the Einstein interaction one gets different theories,
as we show in the following.
\par
In order to make properly this limit, we expand around a Minkowski 
background with a dimensionful coupling constant $\ell$, roughly speaking 
the Planck length. 
The fields $e_{\mu}^m,\,\omega_{\mu}^m$ are the connections of 
the gauge group, and have then dimension $L^{-1}$, 
while the spin--two field 
has dimension $L^{-1/2}$. So we define
\beq
e_m&=&\ell^{-1}E_m=\ell^{-1}(\E0_m+\frac{1}{2}\ell^{1/2}
h_m)=\ell^{-1}\E0_m+\frac{1}{2}\ell^{-1/2}h_m\\
\omega_m&=&\ell^{1/2}\Omega_m\,.
\eeq
Here $E^m_{\mu}$ is the dimensionless vielbein, and $\E0^m_{\mu}$ is 
the constant background vielbein. 
We consider the Minkowski background, that is, 
$\E0^m_{\mu}=\d^m_{\mu}$. The fields $h^m_{\mu}$ have dimension 
$L^{-1/2}$, while the fields $\Omega^m_{\mu}$ have dimension 
$L^{-3/2}$ (the dimension of $\Omega$ is defined in such a way that 
the linearized action does not depend on $\ell$).
The action (\ref{defCSaction}), in terms of these fields, 
becomes
\beq
\label{Iexpanded}
I^{'CS}&=&\int \ll(h_m\wedge d\Omega^m+\e_{mnp}\E0^m\wedge
\Omega^n\wedge\Omega^p+\rr.\nn\\
&&\,\,\,\ll.+\frac{1}{2}\ell^{1/2}\e_{mnp}h^m
\wedge\Omega^n\wedge\Omega^p+\frac{1}{3}\bmu\e_{mnp}\Omega^m\wedge
\Omega^n\wedge\Omega^p\rr)
\eeq
where we have defined 
\be
\bmu\equiv \ell^{3/2}\mu\,.
\ee
The gauge transformations (\ref{gaugeEin}) become (after a rescaling of 
the gauge parameters)
\beq
\label{gaugeexpanded}
\d h^m&=&2\ell^{1/2}\d e^m=
2d\epsilon^m-\e^{mnp}(2\E0_n+\ell^{1/2}h_n)\s_p+2\ell^{1/2}
\e^{mnp}\Omega_n
\epsilon_p+\bmu\e^{mnp}\Omega_n\s_p\nn\\
\d\Omega^m&=&d\s^m+\ell^{1/2}\e^{mnp}\Omega_n\s_p\,.
\eeq
\par
If we take the limit
\be
\ell\rightarrow 0\,,\,\,\,\bmu\rightarrow 0
\ee
of the theory (\ref{Iexpanded}), (\ref{gaugeexpanded}), we have
\be
I_0=\int \ll(h_m\wedge d\Omega^m+\e_{mnp}\E0^m\wedge\Omega^n\wedge
\Omega^p\rr)\,.
\ee
After fixing the Lorentz gauge with the condition 
\be
h_{m\mu}=h_{(m\mu)}
\label{gaugefixing}
\ee
we can solve for the auxiliary field $\Omega$
\be
\Omega_{\mu\nu\a}=\partial_{[\mu}h_{\nu]\a}
\ee
finding the Pauli--Fierz action with the abelian gauge symmetry 
$\d h_{\mu\nu}=2\pa_{(\mu}\epsilon_{\nu)}$.
\par
If we take the limit
\be
\ell{\,\,\rm fixed}\,,\,\,\,\bmu\rightarrow 0
\ee
of the theory (\ref{Iexpanded}), (\ref{gaugeexpanded}), we have
\be
I^{E-H}=\int \ll(h_m\wedge d\Omega^m+\e_{mnp}\E0^m\wedge
\Omega^n\wedge\Omega^p+\frac{1}{2}\ell^{1/2}\e_{mnp}h^m
\wedge\Omega^n\wedge\Omega^p\rr)\,.
\ee
By solving for the auxiliary field $\Omega$ and fixing the Lorentz 
gauge as in (\ref{gaugefixing}), one recovers the Einstein--Hilbert 
action with diffeomorphism invariance.
\par
Let us now consider the limit 
\be
\ell\rightarrow 0\,,\,\,\,\bmu\,\,{\rm fixed}
\ee
of the theory (\ref{Iexpanded}), (\ref{gaugeexpanded}),
\be
\label{Inew}
I^{EX}=\int \ll(\frac{1}{2}
\e_{mnp}h^m d\Omega^{np}+\frac{1}{2}
\e_{mnp}\E0^m\wedge
\Omega^n_{\,\,q}\wedge\Omega^{qp}+\frac{1}{3}\bmu\Omega_p^{\,\,m}\wedge
\Omega_{mn}\wedge\Omega^{np}\rr)
\ee
\beq
\d h^m&=&2d\epsilon^m-2\E0_n\s^{mn}-
\bmu\e^{npq}\Omega_{mn}\s_{pq}\nn\\
\d\Omega_{mn}&=&d\s_{mn}\,,
\label{gaugenew}
\eeq
where we have defined $\Omega^{mn}=\e^{mnp}\Omega_p$, 
$\s^{mn}=\e^{mnp}\s_p$.
In the action (\ref{Inew}) the fields $\Omega^{mn}$ are non--propagating, 
then we can solve for them in terms of the $h^m$. However, we cannot find 
a closed expression for $\Omega^{mn}=\Omega^{mn}(h)$, this can be 
worked out only order by order in $\bmu$. So, the action in second order 
formulation has an infinite number of terms.
\par
This is not a real problem, because the action (\ref{Inew}), 
in first order formulation, is perfectly consistent. It is exactly 
invariant with respect to the gauge transformations (\ref{gaugenew}).
Notice that it is not necessary to add terms at higher order in $\bmu$ to 
the gauge transformations, because the term in $\bmu$ in the action 
depends only 
on $\Omega^{mn}$, and the gauge transformation of $\Omega^{mn}$ doesn't 
contain $\bmu$. 
\par
Let us consider the theory (\ref{Inew}), (\ref{gaugenew}) 
in second order formalism, at first order 
in $\bmu$. The gauge transformations, in components, are
\beq
\d h_{\a\mu}&=&2\pa_{(\a}\epsilon_{\mu)}+2\pa_{[\a}\epsilon_{\mu]}
-2\s_{\a\mu}+\bmu\e^{\b\g\d}\Omega_{\b\a\mu}\s_{\g\d}\nn\\
\d\Omega_{\a\b}&=&d\s_{\a\b}
\eeq
where we have used $\E0^m_{\a}=\d^m_{\a}$ to transform the flat indices
in curved indices. We fix the Lorentz gauge by imposing 
\be
h_{\a\mu}=h_{(\a\mu)}\,,
\label{gaugefixinf2}
\ee
consistency of the (\ref{gaugefixinf2}) implies
\be
\s_{\a\b}=\pa_{[\a}\epsilon_{\b]}+O(\bmu)\,.
\ee
By varying the action with respect to $\Omega^{ab}$ we find
\be
\Omega_{\mu\a}=\pa_{[\mu}h_{\nu]\a}+O(\bmu)\,,
\ee
so we get
\be
I^{EX}=I_0+\bmu
\int d^3x\ll(\frac{1}{3}\e^{\a\b\g}\eta^{\r\s}\pa_{[\r}h_{\m]\a}
\pa^{[\m}h^{\n]}_{\b}\pa_{[\s}h_{\n]\g}\rr)+O(\bmu^2)
\ee
and
\be
\d_{\epsilon} h_{\a\mu}=2\pa_{(\a}\epsilon_{\mu)}+\frac{1}{2}\bmu\e^{\b\g\d}
\ll(\pa_{[\b}h_{\a]\mu}+\pa_{[\b}h_{\mu]\a}\rr)\pa_{\g}\epsilon_{\d}
+O(\bmu^2)\,,
\ee
which is the ''exotic'' deformation (\ref{a0d3}), (\ref{a1d3}) 
found in the previous section, specialized to a single spin--two field.
\par
The gauge algebra, at first order in the deformation parameter, can be 
read off from $a_2$ (\ref{a2d3}):
\be
[\d_{\epsilon},\d_{\eta}]=\d_{\tau}
\ee
with
\be
\tau^{\a}\equiv\frac{\bmu}{8}\e_{\b\g\d}(\pa^{[\a}\epsilon^{\b]}
\pa^{[\g}\eta^{\d]}-\pa^{[\a}\eta^{\b]}
\pa^{[\g}\epsilon^{\d]})+O(\bmu^2)\,.
\ee
We stress again that this algebra is a deformation of the abelian gauge algebra, 
different from the diffeomorphism algebra.
\par
\subsection{Counting the degrees of freedom}
\par
It is a well known fact that gravity in three dimensions has no 
local physical degrees of freedom \footnote{see \cite{witten}
for a discussion on this point.}.
The same holds for the Pauli--Fierz theory. One could ask if the 
''exotic'' theory (\ref{Inew}), (\ref{gaugenew}) 
has or not physical degrees of freedom.
To answer to this question, we perform the canonical analysis of 
that theory
(cfr \cite{BOOK} for the canonical approach to constrained systems).
\par
We start with the free Pauli-Fierz action in three dimensions.
This analysis extends then easily to the case of (\ref{Inew}).
The action is
\be
S^{\rm{C.S.}}_{\rm{free}}=\int d^3x{\cal L}
=\int d^3x \e^{\m\n\r}
\left[ \frac{1}{2} h_{\mu m} \partial_{\nu} 
\Omega^{~m}_{\rho}
+\frac{1}{2}\partial_{\rho}h_{\m m}\O^{~m}_{\n}
+\e_{mnp}\d^m_{\m}\O^{~n}_{\n}\O^{~p}_{\r}\right]\,.
\ee
The configuration space has dimension $9+9=18$, so the phase
space has dimension $36$. There are $18$ primary constraints. 
Consistency of the constraints gives us $6$ secondary 
constraints, which are preserved in time.
Out of these $24$ constraints, $12$ are first class.
The $12$ remaining are then second class and the number
of physical degree of freedom is $0$, as it should.
\par
Adding the new term 
\be
\cl^{exo.}=\frac{1}{3}{\m}\O^{~m}_{\m}\O^{~n}_{\n}
\O^{~p}_{\r}\e_{mnp}\e^{\m\n\r}
\ee
to the lagrangian, the primary constraints don't change.
The canonical hamiltonian acquires a new term which affects 
the calculation of the consistency of
the constraints.
However, even if the expression of the secondary constraints change, 
its number and type (which are first class, which are second class) 
do not change. So, even the theory (\ref{Inew}), (\ref{gaugenew}) 
has no local physical degrees of freedom.
\par
\subsection{The case of $N$ spin--two fields}
\par
The theory (\ref{Inew}), (\ref{gaugenew}) can be simply generalized to 
the case of $N>1$ spin--two fields. It becomes
\beq
\label{multiInew}
I^{EX}&=&\int \ll(\frac{1}{2}\d_{ab}\ll(
\e_{mnp}h^{m\vert a}\wedge d\Omega^{np\vert b}
+\e_{mnp}\E0^m\wedge
\Omega^{n\vert a}_{\,\,q}\wedge\Omega^{qp\vert b}\rr)+\rr.\nn\\
&&\,\,\,\,\ll.+\frac{1}{3}\bmu_{abc}\Omega_p^{\,\,m\vert a}\wedge
\Omega_{mn}^b\wedge\Omega^{np\vert c}\rr)
\eeq
\beq
\d h^{m\vert a}&=&2d\epsilon^{m\vert a}-2\E0_n\s^{mn\vert b}-
\bmu^a_{\,\,bc}\e^{npq}\Omega_{mn}^b\s_{pq}^c\nn\\
\d\Omega_{mn}^a&=&d\s_{mn}^a
\label{multigaugenew}
\eeq
where $\bmu_{abc}\equiv
\d_{ad}\bmu^d_{\,\,bc}$. The action (\ref{multiInew}) 
is invariant under (\ref{multigaugenew}) provided
\be
\bmu_{abc}=\bmu_{(abc)}\,.
\ee
If we solve the equation for $\Omega^a_{mn}$ at first order in $\bmu$, 
we find the Pauli--Fierz action with the deformation and gauge symmetry
(\ref{a0d3}), (\ref{a1d3}).
\par
It is worth noting that because of the gauge invariance 
(\ref{multigaugenew}), which is only at first order in $\bmu$, we know 
that the master equation at higher order has to be satisfied 
without requiring further conditions. So, the result of the previous 
section
\be
\label{a2a2}
(a_2,a_2)=\g c
\ee
can be understood in this context. In particular, while in the case of 
the deformation studied in \cite{multigraviton} 
(\ref{a2a2}) required the associativity of the algebra 
\be
a^a_{b[c}a^b_{d]e}=0\,,
\label{associativity}
\ee
in our case (\ref{associativity}) is not required. As a consequence, the 
proof of the decoupling of the modes given in \cite{multigraviton} is 
no more valid, and the theory  (\ref{multiInew}), (\ref{multigaugenew}) 
actually describes the coupling of a collection of massless spin--two 
fields.
\par
\vskip .5cm
{\leftline{\bf\Large Acknowledgements}}
\vskip .2cm
\par
First of all, we are extremely grateful to Marc Henneaux for 
its invaluable help at several stages of this work.
We thank also G. Barnich, X. Bekaert and C. Schomblond 
for enlightening discussions.
\par
\begin{small}
This work is partially supported
by the ``Actions de Recherche Concert{\'e}es"
of the ``Direction de la Recherche Scientifique - Communaut{\'e}
Fran{\c c}aise de Belgique", by IISN - Belgium (convention
4.4505.86), and by the European
Commission RTN programme HPRN-CT-2000-00131 in which the authors 
are associated to K.\ U.\ Leuven.
The end of this work has been done while participating in the programme 
{\it Supergravity, Superstrings and $M$--Theory} (Sept. 18, 2000 - 
Feb. 9, 2001) of the Centre \'Emile Borel at the Institut 
Henri Poincar\'e 
(UMS 839 - CNRS/UPMC), during which N. B. was supported 
by a Marie Curie fellowship of the European Community programme 
''Improving Human Research Potential and the Socio--Economic Knowledge 
Base'', contract number HPMT-CT-2000-00165, and L. G. by a CNRS support 
obtained by the Centre \'Emile Borel.
\end{small}
\par
\vskip 1cm
{{\bf\LARGE Appendix}}
\appendix
\section{Deformation of $d=3$ gravity}
\par
We know that the Pauli--Fierz action for a single massless spin--two 
field admits two possible deformations: general relativity and the 
deformation (\ref{a0d3}), 
\be
\label{1a0d3}
a_0^{EX}=\frac{1}{3}\e^{\a\b\g}\eta^{\r\s}\pa_{[\r}h_{\m]\a}
\pa^{[\m}h^{\n]}_{\b}
\pa_{[\s}h_{\n]\g}\,.
\ee
One could ask what happens if both of them are turned on. 
What we find is that Pauli--Fierz action (for a single massless spin--two 
field), deformed by general relativity and our ''exotic'' deformation, 
is strictly related to the ''topologically massive gravity'' found by 
S. Deser, R. Jackiw and S. Templeton \cite{DJT}. 
These two theories are formally related by a field redefinition
(more precisely, a BRST transformation), which can be defined only 
perturbatively, order by order in the parameter. 
\par
\subsection{The Deser-Jackiw-Templeton theory}
\par
Let us start by describing the Deser-Jackiw-Templeton (DJT) theory. 
Its action is\footnote{Here we call $\mu$ what in 
the DJT theory is called $1/\mu$.}
\be
\label{defDJTaction}
I^{DJT}\!=\!\int \ll( e_a\!\wedge\!(2d\omega_a(e)+\e_{abc}\omega^b(e)
\wedge\omega^c(e))
-\mu\omega^a(e)\wedge(2d\omega_a(e)+\frac{2}{3}\e_{abc}\omega^b(e)
\wedge\omega^c(e))
\rr)
\ee
where
\be
\label{we}
\omega_a(e)=\frac{1}{2}\e_{abc}\omega^{bc}(e)
\ee
is the solution of the torsion equation $de^a+\omega^{ab}\wedge e_b=0$.
Apparently this second order action looks similar to the first order
action (\ref{defCSaction2}), but in (\ref{defCSaction2})
$\omega$ is a first order propagating 
field, so that action doesn't admit a second order 
formulation. This difference is crucial, because (\ref{defCSaction2}) 
gives (at least at a classical level) general relativity, 
while (\ref{defDJTaction}) does not. 
\par
The DJT theory describes a massive spin--two field, and it is 
invariant under diffeomorphisms. The action (\ref{defDJTaction}) is the 
sum of the Einstein--Hilbert action and of the action of conformal 
gravity in three dimensions. The theory with this latter action alone 
is conformally invariant, and it as been proven \cite{wittenconformal} 
to be equivalent to the Yang--Mills theory of the conformal group in 
three dimensions (with a Chern--Simons action). 
However, in the DJT theory, the former 
Einstein--Hilbert term break conformal invariance. 
\par
The linearization of the DJT action is
\be
\label{DJTlin}
I_0=\int d^3x\ll(-h_{\mu\nu}G_L^{\mu\nu}-\mu\e_{\a\b\mu}\pa^{\a}h^{\b}_{\,
\,\nu}G_L^{\mu\nu}\rr)
\ee
where $G_L^{\mu\nu}$ is the linearized Einstein tensor.
\par
\subsection{Deformation of Pauli--Fierz theory and DJT theory}
\par
Let us consider general relativity, seen as a deformation of the Pauli 
Fierz theory. The solution of the master equation is
\be
W=W_0+\int d^3x 
\ll(\lambda a^{(\lambda)}+\lambda^2 a^{(\lambda^2)}\rr)+\dots
\ee
where
\be
W_0=\int d^3x\ll(-h_{\mu\nu}G^{\mu\nu}+2h^{*\a\b}\pa_{\a}C_{\b}\rr)
\ee
and 
\be
a^{(\lambda)}=a^{(\lambda)}_0+a^{(\lambda)}_1+a^{(\lambda)}_2\,,
\ee
$a^{(\lambda)}_0$ is the cubic vertex of the Einstein--Hilbert action, 
\beq
a^{(\lambda)}_1&=&-2h^{*\a\b}\G_{\a\b\g}C^{\g}\\
a^{(\lambda)}_2&=&C^{*\a}C^{\b}\pa_{\a}C_{\g}\,.
\eeq
This theory is invariant under diffeomorphisms. The gauge symmetry and 
gauge algebra can be read off from the solution of the master equation
by looking at their antibrackets with the field $h_{\a\b}$ and the ghost 
$C^{\a}$, respectively. In our case, the contributions to these 
antibrackets come 
only from the terms with antighost numbers $1$ and $2$ respectively. 
We have
\footnote{Here and thereafter, the antibracket between two local 
quantities $(a(x),b(y))$ is a shorthand notation for $(\int \!a,b(x))$.}
\beq
\label{diffeoGR1}
(2h^{*\a\b}\pa_{\a}C_{\b}+
\lambda a_1^{(\lambda)},h_{\a\b})&=&2\pa_{(\a}C_{\b)}-2\lambda
\G_{\a\b\g}C^{\g}
\\
\label{diffeoGR2}
(\lambda a_2^{(\lambda)},C^{\a})&=&-\lambda C^{\g}\pa_{\a}C_{\g}\,.
\eeq
\par
Let us consider now the DJT theory. Its action, up to order 
$\lambda\mu$, is 
\beq
\label{DJTlmu}
I^{DJT}&=&I_0+\int d^3x\ll(\lambda a_0^{(\lambda)}-
\mu\e_{\a\b\mu}\pa^{\a}h^{\b}_{\,\,\nu}G_L^{\mu\nu}\rr.\nn\\
&&\ll.+\lambda\mu\frac{1}{3}\e^{\a\b\g}\pa_{[\mu}h_{\nu]\a}
\pa^{[\nu}h^{\rho]}_{\,\,\,\b}\pa_{[\rho}h_{\s]\g}\eta^{\mu\s}
+\lambda\mu G^{\mu\nu}_L\ll(\dots\rr)\rr)+O(\lambda^2,\mu^2)
\,.
\eeq
The term of order $\mu$ is $\d$--exact, and can then be interpreted 
as coming from a BRST transformation generated by
\be
\label{bmu}
b^{(\mu)}=h^{*\a\b}\,\,
\frac{1}{2}\e_{\mu\nu(\a}\pa^{\mu}h^{\nu}_{\,\,\b)}\,, 
\ee
which is equivalent to the field redefinition
\be
\label{fieldredef}
h_{\a\b}\longrightarrow h_{\a\b}+\mu
\frac{1}{2}\e_{\mu\nu(\a}\pa^{\mu}h^{\nu}_{\,\,\b)}
\,.
\ee
The term of order $\lambda\mu$ in (\ref{DJTlmu}) is the ''exotic'' 
deformation (\ref{1a0d3}), plus other BRST--exact 
terms. So, up to order $\lambda\mu$, by deforming the 
free Pauli--Fierz action 
with both the general relativity deformation and 
the deformation (\ref{1a0d3}), we get the DJT theory.
\par
It is possible to go deeper into this statement, and prove that 
such a theory is diffeomorphism invariant up to order $\lambda\mu$, 
by using the cohomological tools. We add to general relativity the 
deformation discussed in section 3, with a coupling constant $\lambda\mu$:
\be
W^{GR+EX}=W_0+\int d^3x
\ll(\lambda a^{(\lambda)}-\frac{1}{2}\lambda\mu a^{EX}
\rr)+O(\lambda^2)
\ee
where $a^{EX}=a^{EX}_0+a^{EX}_1+a^{EX}_2$ has been defined in (\ref{a0d3}), 
(\ref{a1d3}), (\ref{a2d3}). Then we make a BRST transformation generated by 
\be
\label{btilde}
\tilde{b}\,:\,\,\,\,f\rightarrow (f,b^{(\mu)}+\lambda b^{(\lambda\mu)})
\ee
where $b^{(\mu)}$ has been defined in (\ref{bmu}), and 
\beq
b^{(\lambda\mu)}&=&-\frac{1}{4}h^{*\a\b}\e^{\mu\nu\g}h_{\nu\b}
\pa_{\mu}h_{\g\a}+\frac{1}{2}h^{*\a\b}\e^{\mu\nu}_{\,\,\,\,\a}
(\frac{1}{4}\pa_{\b}h_{\nu\g}+\pa_{[\nu}h_{\g]\b})h_{\mu}^{\,\,\g})+\nn\\
&&\frac{1}{4}C^{*\b}C^{\g}
\e_{\mu\nu\g}\pa^{\g}h^{\n}_{\,\,\b}
+\frac{1}{8}C^{*\b}\e^{\mu\nu\g}h_{\g\b}\pa_{\mu}C_{\nu}+
\frac{1}{4}\e^{\mu\nu}_{\,\,\,\,\b}h_{\mu}^{\,\,\g}\frac{1}{2}
(\pa_{\nu}C_{\g}+\pa_{[\nu}C_{\g]})\,.\nn\\
\eeq
The solution of the master equation becomes
\beq
\label{ebSgrex}
&&W^{'GR+EX}=e^{\mu\tilde{b}}W^{GR+EX}=\nn\\
&&=W_0+\int d^3x\ll[\mu (W_0,b^{(\mu)})+\lambda a^{(\lambda)}
+\mu\lambda \ll((a^{\lambda},b^{\mu})-\frac{1}{2}a^{EX}\rr)\rr]
+O(\lambda^2,\mu^2)\,.
\eeq
This theory is diffeomorphism invariant up to order $\lambda\mu$. 
In fact, we have
\beq
&&\ll(W^{'GR+EX},h_{\a\b}\rr)\!=2\pa_{(\a}\tilde{C}_{\b)}-2\lambda\G_{\a\b\g}
\tilde{C}^{\g}+O(\lambda^2,\mu^2)\\
&&\left(W^{'GR+EX},\tilde{C}^{\a}\right)=-\lambda\tilde{C}^{\g}\pa_{\a}
\tilde{C}_{\g}+O(\lambda^2,\mu^2)
\eeq
where $\tilde{C}^{\a}$ is the ghost corresponding to the new 
diffeomorphism parameter, defined as
\be
\tilde{C}^{\a}=C^{\a}-\frac{1}{4}\mu\e_{\a\mu\nu}\pa^{\mu}C^{\nu}\,.
\ee
Higher order 
deformations could be necessary to restore diffeomorphism invariance to 
orders higher than $\lambda\mu$. However, by confronting the 
derivative structure of that terms it is simple to see that such possible 
further deformation cannot be of the kind studied in this paper, that is, 
deformations changing the gauge algebra, namely having $a_2\neq 0$. 
\par
It is worth noting that the theory without the deformation $a^{EX}$ 
is simply general relativity, expressed in different fields. 
In terms of this new fields, defined by (\ref{fieldredef}), it 
is no more diffeomorphism invariant, 
because it doesn't satisfy relations with the 
structure of (\ref{diffeoGR1}), (\ref{diffeoGR2}). 
The deformation $a^{EX}$ is exactly what we need in order to restore 
diffeomorphism invariance after the field redefinition.
\par
However, terms in the BRST transformation at higher 
order in $\mu$ are necessary to transform $W^{GR+EX}$ in the DJT theory. 
Then, the field redefinition is polynomial in $\mu$, and this series 
can be defined only order by order in $\mu$. It cannot converge to an 
invertible field redefinition, because nonperturbatively the 
theory $W^{GR+EX}$ has no degrees of freedom, while the DJT theory 
has two degrees of freedom \cite{DJT}: it describes a spin--two field 
with mass $1/\mu$ (in our notations). In general, such a change of degrees 
of freedom can formally occur when the field redefinition involve derivatives.
\par

\end{document}